\documentclass[prd,a4paper,twocolumn,floatfix]{revtex4}
\usepackage{epsfig}
\begin{document}
\title{A prelude to Neutron Stars: The phase diagram of the strong
interactions at finite density}
\author{Vikram Soni}
\email{vsoni@del3.vsnl.net.in}
\affiliation{National Physical Laboratory, K.S. Krishnan Marg, New Delhi 110012, India}
\author{Dipankar Bhattacharya}
\email{dipankar@rri.res.in}
\affiliation{Raman Research Institute, Bangalore 560080, India}
\begin{abstract}
We consider strong interactions at finite density in mean field theory, 
through an effective
lagrangian that can describe both nuclear matter and quark matter. This
lagrangian has three couplings that are all fixed by experiment and no
other parameters. With increasing baryon density we then find the
following hierarchy. At nuclear density and above we have nuclear matter
with chiral spontaneous symmetry breaking (SSB), followed
by the pion condensed quark matter, again with chiral SSB, albeit with a
different realization and finally a transition to the diquark CFL state
which also has chiral SSB  (and colour SSB), with yet another
realization. To one's surprise at zero temperature (in mean field theory), 
at any finite density chiral symmetry is never restored!

We find another remarkable feature and this is that the tree level mass of
the sigma particle, that is set by experiment to about 800 MeV, has a
crucial and unexpected influence on the physics. Strange quark matter
and strange stars are ruled out for a sigma mass above 700 MeV and
neutron stars with magnetic pion condensed cores, that could provide
magnetic fields of neutron stars, exist only for a small interval, between 
$750$--$850$~MeV, for the sigma mass.
\end{abstract}
\maketitle
\section{\label{sec:intro}Introduction}
     Neutron stars have been a subject of abiding interest for several
  decades, almost since Landau suggested their existence, shortly after the
  discovery of the neutron.  There are a variety of astrophysical phenomena 
  that arise from the physics of neutron stars.  The supernova explosion 
  through which the star is born, is a spectacular luminous event.
  Many neutron stars work as pulsars, which generate beamed radiation 
  in their intense magnetic fields. Neutron stars in binary systems may 
  accrete matter from their companions, giving rise to some of the brightest 
  X-ray sources in the sky.  Some neutron stars with super-strong magnetic 
  fields produce occassional strong bursts of gamma rays (Soft Gamma Repeaters).  
  Most of these phenomena require us to understand the physics of matter at 
  very high density, which govern the mass and the size of neutron stars. 
  In other words, one needs to have a clear understanding of the equation of 
  state of the ground state of superdense matter.  Although much effort has 
  gone into this enterprise over the last four decades it still remains poorly
  understood.  Why?

  Central densities of neutron stars are high,  $\sim$5 to 10
  times nuclear density $\rho_{\rm nuc} = 0.17$~fm$^{-3}$.  For a single species,
  neutrons, this naively translates into a fermi gas with typical fermi momentum, 
  $k_f^{\rm N} \sim  700$~MeV. On the other hand nucleons have structure and a 
  typical size of the order of a fermi (200 MeV)$^{-1}$.  
  It is  clear that at such high
  densities nucleons (neutrons) cannot be treated as elementary.  They are
  composite and resolved.  They are colour singlet bound states of three
  valence quarks.  At such high densities, therefore, treating nucleons as point 
  particles interacting via two body (or more) forces will be inadequate.  
  Yet most available equations of state adopt this approach and therefore fail 
  to capture the correct physics.

    On the other hand, if we use quarks as the elementary degrees of freedom,
  we are presently bound by the fact that only perturbative calculations can be 
  done for QCD.  This implies that calculations can be done in QCD only
  at very high density when the theory is approximately in an  Asymptotically
  Free (AF) phase.  However, at intermediate and low density (close to nuclear
  density), where a nucleonic description is valid, we cannot use perturbative
  QCD as the coupling becomes strong and the physics non perturbative.
  This is the dilemma.

       There are attempts to model the physics by a two phase structure -- a 
  quark matter core with a hadronic/nucleonic exterior shell and crust.  
  Since there is no simple way to link the two phases without using separate 
  parameters for both, this description is somewhat arbitrary. Further, the
  nature of the quark matter state is not clear -- for example, is it in a
  spontaneous chiral symmetry broken state.

Can we find a theory that can describe both these domains? We present, here,
an Effective Chiral Intermediate Lagrangian, $L$, that has quarks, gluons and a
chiral multiplet of $[\vec\pi, \sigma ]$ that flavor-couples only to the
quarks \cite{ref1,ref2,ref3,ref4,ref5,ref6}.

        This  yields the nucleon as quark soliton -- a bound state of quarks
 in a solitonic background of scalar/pseudoscalar field  expectation values
 (EV-s) that follow from the spontaneous breaking of chiral symmetry
  \cite{ref1,ref2}. In this way we can generate nucleon matter from these
 nucleons at low density with a transition to quark matter at high density
 but with the same effective Lagrangian covering both domains. Such a unified
 description has not been given before and depends just on the coupling
 constants of the theory and not on any parameters.

To begin with let us consider the two main features of the strong
interactions at low energy. These are i) that quarks  are
confined as hadrons and ii) chiral symmetry is spontaneously
broken (SSB) with the pion as an approximate Goldstone boson.
There is no specific reason that these two phenomena should
occur at an identical temperature scale, though QCD lattice
simulations show that for $ SU_2(L)\times SU_2(R)$ they are close.
The problem in giving an unequivocal answer to this question
is that we are yet to find a solution to the non-perturbative aspects of
QCD.

 Let us consider QCD with a two flavour $SU_2(L)\times SU_2(R)$ chiral
 symmetry. First let us address the question of what occurs at a lower energy 
 scale, confinement or  chiral symmetry breaking. The problem is that though there
 is a bonafide order parameter for chiral symmetry breaking -- the
 mass of the constituent quark, the Wilson loop is no longer an order
 parameter for confinement, in the presence of dynamical quarks.
 However, by looking at the energy density or specific heat we can get a
 fair idea of the change in the number of operational degrees of freedom or
  particle modes. Such lattice calculations indicate that the change from the
 large number of degrees of freedom in the quark matter phase to few degrees
 of freedom in the hadronic one takes place in one broad step in temperature,
 indicating that the two transitions may be close for  $SU_2(L)\times SU_2(R)$ .

Also, if the chiral symmetry restoration (energy/temperature) scale was lower
than the confinement scale we would expect hadrons to show parity
doubling below the confinement scale but above the chiral
SSB  scale. This is not seen in finite temperature lattice
simulations.

Actually, QCD can have multiple scales \cite{ref5}.  
Apart from a confinement scale and a chiral symmetry restoration scale 
we also have a compositeness scale for the pion. The above considerations 
suggest these scales are respectively in ascending order in energy (temperature). 

Let us consider the interacting fermi liquid of nucleons -- nuclear matter. As
the baryon density is raised beyond overlap, we expect a transition to quark
 matter. An interesting question arises: Is the quark matter in a chiral
 SSB state with constituent quarks or is it, as is usually assumed, in a
 chirally restored state with current quarks? As we will see the Lagrangian $L$,
 given below, answers this question \cite{ref6,ref7}.

On the other hand there is some evidence for this intermediate
Chiral Lagrangian that has simultaneously no confinement but
chiral SSB. Such an effective Lagrangian has quarks, gluons and a
chiral multiplet of
 $[\vec\pi ,\sigma ]$ that flavor couples only to the quarks.
\begin{widetext}
\begin{equation} 
L = - \frac{1}{4} G^a_{\mu \nu} G^a_{\mu \nu}
    - \sum {\overline{\psi}} 
      \left(D\!\!\!\!/ + g_y(\sigma + i\gamma_5 \vec \tau.\vec \pi)\right) \psi
    - \frac{1}{2} (\partial_\mu \sigma)^2 
    - \frac{1}{2} (\partial_\mu \vec \pi)^2 
    - \frac{1}{2} \mu^2 (\sigma^2 + \vec \pi^2) 
    - \frac{\lambda^2}{4} (\sigma^2 + \vec \pi^2)^2 + \hbox{const}
\end{equation}
\end{widetext}
The masses of the scalar (psuedoscalar) and fermions follow from the
minimization of the potentials above. This minimization yields
\begin{equation}
\qquad \mu^2 = - \lambda^2 <\sigma>^2
\end{equation}
It follows that
\begin{equation}
\qquad m_{\sigma}^2 = 2\lambda^2<\sigma>^2
\end{equation}
Experimentally, $<\sigma> = f_\pi$, the pion decay constant.
 This theory is an extension of QCD by additionally coupling the quarks
to a chiral multiplet, $( \vec\pi$ and $\sigma )$ \cite{ref1,
ref2,ref3,ref4}.

This Lagrangian has produced some interesting physics at the mean
field level \cite{ref4,ref8}

 (i) It provides a quark soliton model for the nucleon in which the nucleon
 is realized as a soliton with quarks being bound in a skyrmion
 configuration for the chiral field expectation values (EV) \cite{ref1,ref4,ref8}.

 (ii) Such a model gives a natural explanation for the `Proton spin puzzle'.
 This is because the quarks in the background fields are in a spin-isospin
 singlet state in which the quark spin operator averages to zero. On the
 collective quantization of this soliton to give states of good spin and
 isospin the quark spin operator acquires a small non zero contribution
 \cite{ref9}.

 (iii) Such a Lagrangian also seems to naturally produce the Gottfried sum
  rule \cite{ref10}.

 (iv) Such a nucleon can also yield from first principles (but with some
 drastic QCD evolution), structure functions for  the nucleon which is close
  to the experimental ones \cite{ref11}.

 (v) In a finite temperature field theory such an effective Lagrangian also
  yields screening masses that match with those of a finite temperature QCD
  simulation with dynamical quarks \cite{ref12}. This work also does not show 
  any parity doubling for the hadronic states.

 (vi) This Lagrangian also gives a consistent equation of state for strongly
 interacting matter at all density  \cite {ref4,ref7,ref13}.

 This $L$ has a single dimensional parameter, $f_\pi $, that  is the pion decay
 constant, and three couplings, $g_3$, the QCD coupling, $g_y$, the Yukawa coupling
 between quarks and mesons, that will be determined from the nucleon mass and
 the meson-meson coupling,  $\lambda$, which, for this model, can be determined 
 from meson meson scattering \cite{ref14}. No further
 phenomenological input will be used. As it stands, there is no confinement
 in this model, but may be dynamically generated as in QCD.

 Since this is posited as an effective Lagrangian, we should have an appoximate 
 idea of its range of validity. We find, somewhat in analogy with the top quark 
 (large Yukawa coupling) composite higgs picture, that we can get a compositeness
 scale for the scalars in this model by using Renormalisation Group (RNG) 
 evolution. We find that the wavefunction
 renormalisation for the scalars is inversely proportional to the running
 Yukawa coupling and thus naively vanishes when the Yukawa coupling blows up.
 For our theory such a ballpark scale falls between 700--800 MeV
 \cite{ref15}.

 An independent and quite general approach in setting a limit to the range
 of validity of non
 asmptotically free (e.g. Yukawa) theories is the vacuum instability
 to small length scale fluctuations (or large momenta in quantum loop corrections) 
 that plagues
 these theories, discovered and analytically proved by one of us \cite{ref16}.
 The scale at which this occurs is of the same order as above. This is not
 very surprising since it is connected to non-AF character of the Yukawa
 coupling \cite{ref16,ref17}. This underscores the impossibility of doing loop 
 calculations, unless we introduce a cut off, even though our $L$ is renormalizable!

 Given these facts we shall use this theory  at the Mean Field level to look
 at different phases of this field theory.  To do this we must first
 establish the ground state of the Baryon Number B=1 sector of this theory, 
 i.e. the nucleon.

 The plan of the paper is as follows,
 In Sec.~2 we review the description of the nucleon in this model, which 
 fixes $g_y$.
 This is followed in Sec.~3 by a pedagogical description of how we may look 
 upon nuclear matter
 at density above nucleon overlap. We then look at some phases of 2 flavour 
 quark matter
 in Sec.~4 and point out that the phase with lowest ground state energy is the pion
 condensed phase with chiral SSB. In Sec.~5 these results are generalized to 3
 flavours. In Sec.~6 we review and  compare the pion condensed phase with the
 diquark condensed colour supercondcting phase.
 We  end this section with a review of the phase diagram of QCD at all density. 
 In Sec.~7 we remark on the phases to be used in constructing neutron stars in
 a following paper and discuss the validity of Mean Field Theory (MFT).

 \section{The Nucleon in the Chiral Linear Sigma Model with Quarks}
  The basic fields are the three component (isospin) pion fields, $\vec\pi(r)$, 
  the scalar field, $\sigma(r)$, which together form a real 4 component
  chiral multiplet that transforms as a representation of O(4), the rotation
  group in four dimensions.  The fermionic fields are the quarks 
  which transform as a fundamental representation $SU_L(L) \times SU_R(R)$
  which is isomorphic to O(4). In this we ignore the gluon fields which are 
  vector and therefore not expected to carry expectation values in MFT. 
  However, corrections due to one-gluon exchange interactions between quarks 
  can be included, but we shall not do so below, as their effect is small. 
  We also neglect the pion mass corrections. The Hamiltonian
  \cite{ref1} which is invariant under the above group transformations then reads
\begin{widetext}
   \begin{equation}
    H =  \int d^3x \left[\frac{1}{2}(\partial_i\vec\pi)^2 +
         \frac{1}{2} (\partial_i\sigma)^2 + V(\sigma,\vec\pi) 
       + \sum {\psi^{\dagger} \left( -i\partial_i\alpha_i + \beta g_y(\sigma +
        i\gamma_5 \vec \tau .\vec \pi)\right) \psi}\right]
       + O(m_\pi)
     \end{equation}
\end{widetext}
  The term $V (\sigma,\vec\pi)= \frac{\lambda^2}{4}(\sigma^2 +\vec\pi^2 -f_{\pi}^2)^2$
   is the potential functional.  For the vacuum  sector (no fermions)
  this is the
  quantity to be minimized.
  The minimum of $V (\sigma,\vec \pi) $ occurs at $\sigma^2 + \vec\pi^2 =
  f_{\pi}^2$.
  This is the equation of a 3 sphere and thus allows for a continuous 
  degeneracy. However, normally the choice, $<\sigma> = f_{\pi}, <\vec\pi> = 0
  $ is
  made, which is consistent with the pseudoscalar, $<\vec\pi>$, having a zero
  Vacuum Expectation Value (VEV) to avoid spontaneously violating parity in 
  hadronic scattering. Also,
  with this choice the goldstone pseudoscalar excitations about the ground state
  are the right ones -- the psuedoscalar pions.
  Once this choice for the vacuum state is made, it is clear that the vacuum
  state spontaneously violates the chiral symmetry, it changes under a O(4)
  rotation, even though $H$ does not.

  We now move to the description of the ground state for the 
  the baryon number B = 1 sector.

  1) The usual pattern of symmetry breaking is the space uniform VEV
  corresponding to the choice above.
     \begin {equation}
  <\sigma> = f_{\pi} , <\vec\pi> = 0
      \end {equation}
   This choice gives a spontaneous or Yukawa mass to the quark,
   $m_q = g<\sigma>$.

   The lowest energy for B = 1 sector, which corresponds to the quantum 
   numbers of the nucleon, is $M = 3 m_q = 3 gf_{\pi}$ -- simply the mass 
   of three quarks.

   2) Let us now consider the case when a fermion bound state can arise from
   a time independent expectation value (EV) that is locally space dependent.  
   Clearly, to have finite energy for this state requires that asymptotically 
   (far away from the localized fermion source), the EV revert back to the VEV 
   above.
   We now move on to the so called skyrme hedgehog configuration,
        \begin {equation}
       <\sigma> = f_{\pi}\cos{\theta(r)},\;\;\;  
       <\vec\pi> = \hat r f_{\pi}\sin{\theta(r)}  
        \end {equation}
   where $\theta(r\rightarrow\infty) = 0$,  from the finite energy condition
and $\theta(r\rightarrow 0)= -\pi$,  for the pion field to be well defined
     at the origin.

 It may be pointed out that we have chosen  $\sigma^2 + \vec\pi^2 =
 f_{\pi}^2$. More generally, we need not fix this magnitude and vary
 $\sigma$ and $\vec\pi$ independently.
 Given such a configuration, we solve the Dirac eigenvalue equation for
 the quark in the background field, $\theta(r)$.  Due to the linking of space
 with internal isospace, neither the angular momentum, $\vec J$,
 nor the
 isospin, $\vec I$, commute with the Dirac Hamiltonian but only the sum
 $\vec K = \vec J +\vec I$ does.
 $K$ can then be used to label the eigenstates.  The lowest,
 $K = 0$, valence state is a spin isopin singlet of the form
 \begin{equation}
    \psi_{K=0}  =  \psi_0(r) |K=0\rangle
 \end{equation}
 where,
 \begin{equation}
    |K=0\rangle = \frac{1}{\sqrt2}\left(|\!+\!1/2\rangle|\!\uparrow\rangle 
                  \; - \; |\!-\!1/2\rangle|\downarrow\rangle\right)
 \end{equation}
 the arrows designate spin and the halves the isospin.
 The eigenvalue equation is
 \begin {equation}
      (-i\partial_i\alpha_i + \beta g_y(\sigma +
 i\gamma_5 \vec \tau. \vec \pi)) \psi_{K=0}  = \epsilon_0 \psi_{K=0}
       \end {equation}
 The $|K=0\rangle$ state is a bound state, where $R$ has the interpretation of the
    width of the potential.
    For a particular profile, $\theta(r) = -\pi (1-r/R)$, for $r < R$,  
    and $ \theta(r)=  0 $, for $r>R$, the eigenvalue
    dependence on $R$ is sketched in \cite{ref1}(a),\cite{ref4}.

    The nucleon may then be obtained as a bound state of three coloured quarks.
    Notice that the only degeneracy of the $K=0$ state is in colour,
    since this state is a spin isospin singlet and in making the nucleon we
    have exhausted this degeneracy, yielding a colour singlet.  The energy of
    the nucleon state is given by
     \begin{equation}
        E^{{\rm B}=1}[\theta(r)] = 3 \epsilon_0 [\theta(r)] +
      \int d^3x \frac{1}{2}[(\partial_i \vec\pi)^2 + (\partial_i \sigma)^2]
     \end {equation}
    in terms of a general profile, $\theta(r)$.
    This must be minimized with respect to variations of $\theta(r)$ to get the
    ground state energy, $E_{\rm min}$.
    Finally, since we have eigenstates of $K$, this quark soliton must be
    projected into good states of spin and isospin.
    The mass, $M$, of this soliton with quark bound states which has the
    quantum numbers of the nucleon depends just on $f_{\pi}$, $g_y$ and $\lambda$.
    The dependence on $\lambda$ is marginal, so the only parameter that is
    free is the Yukawa coupling, $g_y$. We fix, $M=M_{\rm nucleon}$.

    Actually, for such a nucleon, which is
    a colour singlet bound state of three valence quarks in a skyrme background,
    in a linear sigma model, a generally accepted value for  $g_y =5.4$
    \cite{ref1}(b). This is when the $\pi$ and $\sigma$ fields are varied 
    independently, without making any ansatz,
    and the quark soliton so obtained is projected to give a nucleon with good 
    spin and isospin.

    Let us now compare this to the energy of 3 free quarks in the uniform
    phase.  We find that,
    \[ M/(3 g_y f_{\pi}) <  1  \]
    indicating that in our sigma model the nucleon is indeed a soliton with
    quark bound states. This analysis makes a further significant point.  The
    threshold value of $g_y$ at which $M/(3 g_y f_{\pi})$ is first less than 1,
    is $g_y = 4$.  Therefore, for values of $g_y$ greater than 4, the quark soliton
    is always of lower energy than three free quarks and is thus bound. Besides
    the mass of this nucleon falls with increasing $g_y$, as this controls  
    the strength of the
    attractive potential in which the quarks bind.  The conclusion, therefore, is
    that there is a maximum mass \cite{ref18} for the fermion in such a theory.

     Here we remark that we now have a determination for all parameters of our $L$:
     $f_\pi$, $g_3$ and $g_y$, leaving only one parameter, $\lambda$ to be set.
     As we shall show  this can be determined from low energy meson-meson
     scattering data.

     Recently, Schechter et al \cite{ref14} made a fit to scalar channel 
     scattering data
     to see how it may be fitted with increasing $\sqrt{s}$ (centre of mass energy),
     using chiral perturbation theory and several resonances. They further looked 
     at this
     channel using just a linear sigma model. Their results indicate that for
     $\sqrt{s} < 800$~MeV, a reasonable fit to the data can be made using the
     linear sigma model with a tree level sigma mass above but close to 800 MeV. 
     This sets the value of $\lambda$.

     This completes the  determination of all the parameters of our $L$,
     which is able to describe both nucleon and quark phases of dense matter.

\section{Phases at finite density}
   We now move to the main theme of this work, which is to look at ground state
   of strongly interacting matter at finite density. In our model all the phases
   considered are characterised simply by different patterns of spontaneous
   symmetry breaking. As we shall see i) the nucleonic phase is characterised
   by the pattern of SSB in skyrmions ii) the Lee-Wick phase has space uniform
   SSB, with only the sigma field expectation value and iii) the pion
   condensed phase is characterised by a stationary wave, with a space dependent 
   periodic variation in the sigma and neutral pion field expectation values.

   \subsection{ The nucleon or nuclear phase}
   In this phase, as the name suggests, the quarks are to be found as bound 
   states in nucleons.

   We found that the single isolated solitonic nucleon
   has lower energy than the three free valence quarks. Thus at low density
   we have a fermi gas of interacting nucleons. Long range potentials like pion 
   exchange and
   tensor exchange can be found by the appropriate two nucleon ansatz, in analogy to
   the two-skyrmion problem. These match well with the  usual nuclear physics
   two-nucleon potentials \cite{ref19}. It is then more reasonable 
   to solve the nucleon many body
   problem by mapping on to the nuclear many body calculation -- since it is very
   difficult to solve a many body quark soliton problem!

   We know quite well the ground state of nuclear matter till, say, roughly
   two times
   nuclear density. From there on we can only model this phase at higher
   density, when the solitons start to overlap and their motion is obstructed, 
   as a `crystal lattice' of solitons \cite{ref13}. We can then use the 
   Wigner-Seitz
   approximation to convert the problem to a single cell problem.  Scaling
   the size of the single cell scales the baryon density, since we have one
   nucleon per cell.

   It is appropriate to note that with such simplifying approximations it
   is not fair to expect a realistic description but a pedagogically useful one.

   \vspace*{1ex}
   \noindent
   \textit{The Wigner Seitz approximation}\\
   The crystal ansatz implies a picture where the solitons sit in a close
   packed hexagonal configuration. Each soliton is in a skyrmion chiral symmetry
   broken configuration for the $ [\sigma,\vec \pi] $ fields that acts as a
   potential that supports a single bound state with a colour degeneracy of 3. 
   The $|K=0\rangle$ bound state is saturated for each soliton making
   the soliton into a colour singlet baryon. It is clear that when the solitons
   are close packed, the quark wavefunction will leak out so as to minimize 
   the energy.  From the Kronig-Penny model it is expected that bound states 
   will form a band with the number of states in the band being $3N$,
   where $N$ is the number of baryons stacked together. It is further known that
   the band will splay evenly about the single bound state energy, with the 
   bottom of the band below and the top above the single soliton $(\vec K =
   0)$ bound state energy. Due to the saturation of the bound state, the 
   band will be completely occupied and  a good approximation to the 
   median energy of the band is the  bound state energy of the single soliton
   above of radius $R$.
   Thus the energy per baryon can be approximated in the crystal configuration
   by the single soliton energy but with the parameter $R$ setting the volume
   occupied by a single soliton. Clearly, the baryon density is set by $R$.
        \begin {equation}
                    n_b = \frac{6}{R^3} 
        \end {equation}
   At finite density the VEV becomes an EV and we can allow for
      \begin {equation}
       <\sigma>^2 + <\pi>^2 = F^2
      \end {equation}
   where $F$ is now a variational parameter to be set by minimizing the
   free energy. It is clear that this ansatz is rather restrictive.

Note that at finite density there is just one constraint, namely 
\begin{enumerate}
  \item[i)] $ <\vec\pi> $ must vanish at the origin of the soliton for the pion
         field to be well defined.
\end{enumerate}
 However, only for calculational simplicity we also maintain ii) and iii), below:
\begin{enumerate}
  \item[ii)]  $F$ is space independent
  \item[iii)] $<\sigma> = -F$ at $r=0$,  or \\
              $<\sigma> = F$  at $r=R$,  or \\
              $<\vec\pi> = 0$ at $r=R$
\end{enumerate}

 We choose a simple profile function, $\theta(r)= \pi (1-\frac{r}{R})$,
   where $R$ is the half-length of the cell.
  
   We may now calculate the Wigner-Seitz single soliton energy 
   by calculating the quark energy eigenvalue $\epsilon_0$, meson gradient
   energy (second term in eq.~(\ref{eq:crystEb})) and symmetry (condensation) 
   energy (third term in eq.~(\ref{eq:crystEb})) in this background: 
   \begin{equation}
      E_b = 3\epsilon_0(\theta(R))
            +2\pi F^2 R \left(1 + \frac{\pi^2}{3} \right)
            +\frac{1}{3}\pi\lambda^2 (F^2 - f_{\pi}^2)^2 R^3
   \label{eq:crystEb}
   \end{equation}
    The additional
    complexity is that we must now minimize $E_b$ for the single soliton
    energy with respect to $F$ for each $R$ or at each density.

    Further relaxing of constraints or solving the full set of Euler-Lagrange
    equations for each field is somewhat laborious, so this is as far as we go. 
    This is so because this Wigner-Seitz approximation has other uncertainties. 
    We refer 
    the reader to a more sophisticated treatment of this problem \cite{ref20}.

    This section, as stated earlier, is to be seen as pedagogical. It does
    not yield a realistic equation of state.

    Clearly the energy eigenvalue goes up as $R$ (width of the potential) is
    decreased, till, at some lowest value of $R$, a bound state solution cannot
    be supported and this phase is lost.

\vspace*{1ex}
\noindent
\textit{Remarks}\\
\begin{itemize}
\item[i)] The Equation of State (EOS) with the above variational ansatz is 
    obviously inadequate to 
    produce the single soliton/nucleon with arbitrary variations of
    $<\sigma>$ and $<\pi>$.  The exact solution fits the nucleon for a value of 
    $g_y=5.4$. This will not be the case here. We will need a larger
    coupling, $g_y$, to fit the nucleon in this approximation.  Clearly, the
    minimum of $E_b$ vs $R$, at $R_{\rm min}$, is the energy of a single 
    isolated soliton.
\item[ii)] The nucleon may be obtained by projection of the soliton into good
    spin isospin states, $ \vec J = \vec I = \frac{1}{2} $.  The soliton is
    largely a wave packet, a linear superposition of all $ \vec J = \vec I =
    (n + \frac{1}{2})$ states.  The maximum weight comes from the lowest, 
    $ \vec J = \vec I $ states.  We make the approximation that it is an equal
    linear superposition of the Nucleon (N) and $\Delta$ states and set the soliton 
    energy to be midway between $M_{\rm N}$ and $M_{\Delta}$.
    \begin {equation}
    E_{b,{\rm min}}=M_{\rm soliton}=M_{\rm N}+\frac{1}{2}(M_{\Delta} -M_{\rm N}) 
    \end {equation}
\item[iii)] Even with this approximation we find that to fit $E_{b,{\rm min}}$, so 
    as to produce a nucleon with $M_{\rm N} = 940$~MeV requires $g_y \sim 6.3$.
\item[iv)] Furthermore, there is a zero point quantum mechanical energy of 
    localization associated with the localized nucleon states that make up the
    crystal that may be estimated from the Uncertainty Principle.
\item[v)] Another correction is that due to one gluon exchange interaction between 
    the quarks.
\end{itemize}
    We leave out the last two corrections and others as we shall be using a
    more realistic EOS anyhow (see below).

    Figure 1 shows the energy per baryon, $E_b$ versus the baryon density
    $n_b$ in this phase.  In this, the minimization of 
    $E_b$ with respect to the variational parameter $F$ has been carried out. 

    As we have pointed out, to construct the exact ground state of the crystal 
    is, to say
    the least, a formidable exercise.  We have tried to employ an educated 
    variational ansatz in the hope that it provides a good approximation.
    The EOS this yields has many correct qualitative features but nevertheless
    is much stiffer than most known EOS -- it may not be unfair to say that it 
    is almost arthiritic.  In this instance it may be more 
    judicious to use a well worn nuclear equation of state, like the APR98 EOS
    \cite{ref21}, for the entire nuclear phase, provided the density
    at which the transition to quark matter takes place is not much above
    twice nuclear density.  A comparison between our 
    stiff EOS and the APR98 is provided in Figure (1).

    The net result is that the nucleon is a soliton, the nucleons interact 
    with each other to produce binding at around nuclear density. We then
    expect to have a Fermi liquid of interacting nucleons till the nucleons
    begin to overlap. As the density increases the $E_b$ increases due 
    to topological repulsion, we expect that the nucleons are no longer free 
    to move around and get localized into a crystal (of nucleons) like 
    configuration analysed above. This solitonic phase then dissolves into  
    quark matter.
%   a chirally restored phase of free massless quarks.
\begin{figure}
\epsfig{file=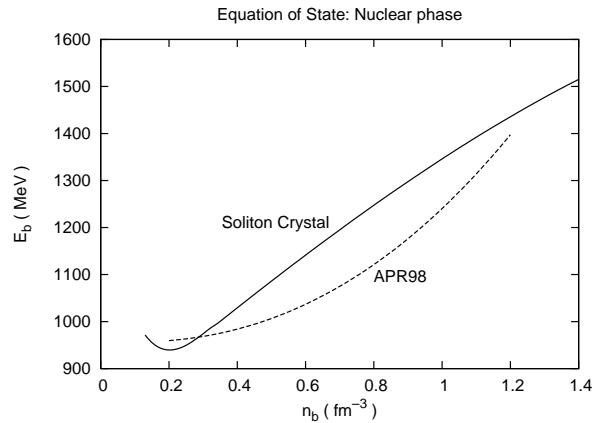,width=80mm}
\caption{Energy per baryon as a function of baryon density in the Wigner-Seitz
    quark soliton crystal as a model of nuclear state (solid line).  The 
    dashed line shows the corresponding relation in the APR98 \cite{ref21}
    equation of state for beta-stable matter, obtained using A18+$\delta v$+UIX 
    interaction model.}
\end{figure}

\section{Two Flavour Quark Matter}

  We shall now consider in Mean Field Theory the phases of two-flavour quark
  matter in the $SU(2)_L \times SU(2)_R$ chiral model above.

  We shall then extend the model to three flavours (u, d, s).

  \subsection{The space uniform phase}

    We now turn to the phase in which the pattern of symmetry breaking is 
    such that the expectation values of the meson fields are uniform. At zero
    density they are just the VEVs. 
    \begin{eqnarray}
                 <\sigma> &=& f_{\pi} \\
                 <\vec \pi> &=& 0 
    \end{eqnarray}
    For arbitrary density we allow the expectation value to change in
    magnitude, as it becomes a variational parameter that is determined
    by energy minimization at each density.
    \begin{eqnarray}
                <\sigma> &=& F \\
                <\vec\pi> &=& 0
    \end{eqnarray}
    Such a pattern of symmetry breaking simply provides a constituent mass to the 
    quark $m = g_y<\sigma> = g_yF$ and the quarks are in plane wave states
    as opposed to the bound states in the nucleonic phase \cite{ref13}.

    The mean field description of this phase is simple. The energy density is
    \begin{equation}
               \epsilon_{\rho} = \sum_{u,d} \frac{1}{(2\pi)^3} \gamma
               \int d^3k \sqrt{m^2 + k^2}  + \frac{\lambda^2}{4} (<\sigma^2>
               -f_{\pi}^2)^2
    \end{equation}
    where $ m = g_y<\sigma> = g_yF$ and the degeneracy $\gamma = 6$. We shall
    use $ g_y = 5.4$ as determined from fixing the nucleon mass in this model 
    at 938 MeV \cite{ref5,ref6}.  The integral above runs up to the `u' and `d' 
    fermi momenta.

    For neutron matter (without $\beta$ equilibrium) we have the relations 
    \begin{eqnarray}
        k^f_u &=& (\pi^2 n_u)^{\frac{1}{3}} = (\pi^2 n_b)^{\frac{1}{3}} \\
        k^f_d &=& (2 \pi^2 n_b)^{\frac{1}{3}} \\
        E_b  &=& \frac{\epsilon_{\rho}}{n_b} 
    \end{eqnarray}
    where $n_b$ is the baryon density.

    At any density the ground state follows from minimising free energy, 
    with respect to $<\sigma> = F$.
    As shown in the figures of \cite{ref4,ref7,ref13}, this phase begins at
    $n_b=0$, with $E_b = 3gf_\pi$, which then falls till 
    chiral restoration occurs at some $n_X$. After this, as the density is 
    increased, $E_b$ continues to drop, goes to a minimum and 
    then starts rising corresponding to a massless quark fermi gas.

   In the chirally restored phase the EOS is very simple and parallels the
   MIT bag description of \cite{ref22}:
   \begin{eqnarray}
     n_b &>& n_X\\ 
     \epsilon_{\rho}
            &=& \left(\frac{3}{4\pi^2}\right) {\pi^2 n_b}^\frac{4}{3}\alpha
                          + \frac{\lambda^2}{4} f_{\pi}^4 
   \end{eqnarray}
   The last term above is just the bag energy density, and
   \begin{equation}
       \alpha = (1 + 2^{\frac{4}{3}})  
   \end{equation}
   This phase has two features, a) chiral restoration at $n_X$ followed,
   with increasing density, by b) an absolute minimum in $E_b$, at a
   $n_C > n_X$.

   From the comparison of this phase with the nucleon and nucleonic `phase'
   arising from the same model (see \cite{ref4,ref13}), it is clear that 
   the nucleonic phase is always of lower energy than the uniform phase 
   above, upto a density of roughly 3 times nuclear density,
   which is above the chiral restoration density in the uniform phase.

      \subsection{The Pion Condensed phase}

   Here we shall consider another realization
   of the expectation value of  $<\sigma>$ and $<\vec\pi>$ corresponding
   to pion condensation. This phenomenon was first considered in the context
   of nuclear matter.

   Such a phenomenon also occurs with our quark based chiral $\sigma$
   model and was first considered at the Mean Field Level by Kutschera
   and Broniowski in an important paper \cite{ref7}. Working in the chiral 
   limit they found the pion condensed state has lower energy than the uniform, 
   symmetry breaking state (phase 2) we have just considered, at densities of 
   interest.
   This is expected as the ansatz for the PC phase is more general than for 
   phase 2.

   The expectation values now carry a particular space dependence
   \begin{eqnarray}
      <\sigma> &=& F  \cos{(\vec q. \vec r)} \\
      <\pi_3>  &=& F  \sin{(\vec q. \vec r)} \\
      <\pi_1>  &=& 0 \\
      <\pi_2>  &=& 0
   \end{eqnarray}
   Note that when $|\vec q|$ goes to zero, we recover the uniform phase (2).
   The Dirac Equation in this background is solved in \cite {ref13}
   and reduces to
     \begin{equation}
     H \chi(k) = (\vec \alpha . \vec k - \frac{1}{2} \vec q . \vec \alpha \gamma_5 
       \tau_3 + \beta m) \chi (k) = E(k)\chi(k)   
      \end{equation}
   where $m=g_yF$.
   The interaction term has been recast in terms of the relativistic spin
   operator, $\vec \alpha \gamma_5$.  It is evident that if spin is parallel 
   to $ \vec q$ and $\tau_3 = +1$ (up quark) then this term is negative and 
   if $\tau_3 = -1$ (down quark) then it is positive.
   For spin antiparallel to $\vec q$ the signs for $\tau_3 = +1$ and $-1$  are
   reversed.

   The spectrum for the hamiltonian is the quasi particle spectrum and
   can be found to be
   \begin{eqnarray}
    E_{(-)}(k) &=& \sqrt{m^2 + k^2 +\frac{1}{4}q^2 -\sqrt{m^2 q^2 +
                          (\vec q.\vec k)^2}} \\
    E_{(+)}(k) &=& \sqrt{m^2 + k^2 + \frac{1}{4}q^2 + \sqrt{m^2q^2 +
                           (\vec q.\vec k)^2}} 
   \end{eqnarray}
   The lower energy eigenvalue $E_{(-)}$ has spin along $\vec q$ for
   $\tau_3=1$, or has spin  opposite to $\vec q$ for $\tau_3 =-1$. 
   The higher energy eigenvalue
   $E_{(+)}$ has spin along $\vec q$ and $\tau_3 = -1$, or has spin
   opposite to $\vec q $ and $ \tau_3 = +1$.

   In  this  background the fermi sea is spin
   polarized into the  states above. The quasi particles are,
   however, good states of $\tau_3$.

   First we fill up all the  lower energy, $E_{(-)}(k)$, states and
   then we have a gap and start filling up the  $E_{(+)}(k)$ states
   till we get to $E_{F}^i$, the fermi energy corresponding to a given
   density for each flavour.
   \begin{eqnarray}
     n_{i} &=& \frac{1}{(2\pi)^3}\gamma \left(\int d^3k\Theta( E_F^i- E_{(-)}(k)) 
	                                    \right.        \nonumber \\
           & & \left. +\int d^3k \Theta( E_F^i- E_{(+)}(k))\right) \\
     n_b &=& (n_u + n_d)/3 \\
     \epsilon_i &=& \frac{1}{(2\pi)^3}\gamma \left(\int d^3k E_{(-)}(k)
                  \Theta( E_F^i- E_{(-)}(k)) \right. \nonumber \\ 
            & & \left. +\int d^3k E_{(+)}(k) \Theta( E_F^i- E_{(+)}(k))\right) \\
     \epsilon_\rho &=& \epsilon_u  +  \epsilon_d + \frac{1}{2} F^2 q^2 + 
                  \frac{\lambda^2}{4}(F^2 - f_{\pi}^2)^2
   \end{eqnarray}
   We can now write down the equation of state as in \cite{ref7}.  It is found 
   that the PC state is lower in energy than the uniform phase 2 for 
   densities of interest.
   For the explicit numbers and figures we refer the reader to \cite{ref7}.

   We briefly remark on some features of this phase. 
   \begin{enumerate}
   \item The reason that the PC phase has energy lower than the uniform $<\sigma>$
      condensate is perhaps best understood in the language of quarks and 
      anti quarks. To make a condenste a quark and antiquark must make a
      bound state and condense.  For a uniform $<\sigma>$ condensate the $q$ and 
      $\bar{q}$ must have equal and opposite momentum. Therefore, as the quark 
      density goes up the system can only couple a quark with $k > k_f$ and 
      a $\bar{q}$ with the opposite momentum. This costs much energy so the
      condensate can only occur if $k_f$ is small, at low density.                                    
      On the other hand, the pion condensed state is not uniform. So at finite
      density, if we take a quark with $k = k_f$ the $\bar{q}$ can have momentum 
      $k =  |\vec k_f -\vec q |$, which is a much smaller energy cost
   \item Since the pion condensate is a chirally broken phase,
      the chiral restoration shifts from very low density in the uniform phase
      to very high density, $\sim 10 \rho_{\rm nuc}$.  This is a signature of 
      this phase.
   \item Since this phase is lower in energy than the uniform phase for all
      densities of interest
      we go directly from the nucleonic phase to the PC phase completely 
      bypassing the uniform phase showing that all the `interesting' features 
      and conjectures for the uniform phase are never realized.
   \item Another feature of this $\pi_0$ condensate is that since we have a spin
      isospin polarization we can get a net magnetic moment in the ground state, as
      the magnetic moments of the u and d quarks add.
   \end{enumerate}

  \section{The three flavour state}

The extension to three flavours or $SU(3)$ chiral symmetry needs some clarification.

     The generalized Dirac Equation for the $SU(3)$ case is considerably more
     complicated and involves a singlet $\xi_0$ and an $SU(3)$ octet $\xi_a$
     of scalar fields and a singlet $\phi_0$ and an $SU(3)$ octet $\phi_a$
     of pseudoscalar fields,
     that interact with the quarks as shown in \cite{ref23}.
\begin{widetext}
     \begin{equation}
     H \psi(k) = (-i\vec \alpha .\vec \partial -
      g_y \beta(\sqrt{2/3}( \xi_0 +i\phi_0 \gamma_5) +
      \lambda^a( \xi_a +i\phi_a \gamma_5))) \psi  =  E\psi
     \end{equation}
\end{widetext}
In the chiral limit, the spontaneous symmetry breaking pattern is not unique.
We choose the pattern in which the $SU(3)_L \times SU(3)_R$ chiral
symmetry breaks down to a vector $SU(3)$. For the uniform case, we have
    \begin{eqnarray}
        <\xi_0>&=& \sqrt{3/2} f_{\pi}\\
                       <\xi_a> &=& 0 \\
                <\phi_0> &=& 0  \\
                     <\phi_a> &=& 0 
    \end{eqnarray}
   This gives a constituent mass  $ m = g_yf_\pi $ for all (u, d and s)
   quarks. The explicit symmetry breaking strange quark mass term with current
   mass $m_s$, is then added to $H$. The strange quark mass, $M_s$, then, turns
   out to be the sum of the constituent and  explicit mass,
   $M_s = g_yf_\pi + m_s$

   Here we run into a puzzle of sorts. The mass term now has two components,
   an explicit or current mass, $m_s$ and a constituent mass,
   $g_y<\sigma> =  g_yf_\pi$. However, it does not disturb any relation
   (e.g. the GT relation), whether we choose,  $<\sigma> = +$ or $- f_\pi$.
   This raises an ambiguity about the relative sign of the two mass terms.
   It would seem that the choice of opposite signs for the two terms is optimal
   since it gives the lowest mass or ground state energy for the one fermion
   sector.

   When the strange quark gets a large explicit mass $m_s$ about
   150 MeV (the u and d quarks have negligible explicit masses),
   this question becomes very relevant.  However, experimentally the 
   strange baryons have larger masses than the non strange ones suggesting
   that the relative sign is plus. Georgi, in his book 
   \cite{ref24}, finds that the success of the non relativistic quark
   model, particularly for the masses and the magnetic moments of the baryons,
   follows from
   taking constituent masses of about 350 MeV for the u and d quarks and a 
   total mass of about 550 MeV for the strange quark. Clearly, the larger 
   mass of strange baryons suggests the same. Thus, experimentally, the 
   relative sign, plus, is selected.  We shall therefore continue to use this.

 \subsection{The three-flavour Pion Condensed phase}

 For describing strange quark matter we use the 3-flavour Pion Condensed state
 \cite{ref6}.
 This is  a more versatile state than the one used in \cite{ref22}
 (3 flavour Chirally Restored Quark Matter -- CRQM), the latter being a 
 subset of the former.

 Next, we formulate the symmetry breaking in the presence
 of the pion condensate. This is given as follows,
 \begin{eqnarray}
     <\xi_0> &=& \sqrt{3/2} F(1 + 2\cos{(\vec q.\vec r)})/3 \\
     <\xi_8> &=&-\sqrt{3} F(1 - \cos{(\vec q.\vec r)})/3   \\
     <\phi_0>&=& 0   \\
     <\phi_3>&=&  F (\sin{(\vec q.\vec r)})   
 \end{eqnarray}
 while all other fields have zero expectation value.

 This gives exactly the PC hamiltonian equation for the u, d sector
 and continues to give the simple mass relation for the strange quark,
 $ M_s = g_yF + m_s$.
 When $q=0$ and $m_s=0$ we recover the chiral limit above.

 We may now simply add the two-flavour PC results for the
 energy density and density derived above to the strange quark
 energy density which arises from the single particle relation,
 \[E_s = \sqrt{ M_s^2 + k^2}\]
 The strange quark energy density is given by Baym (eqn. 8.20) \cite{ref25}
 \begin{equation}
    \epsilon_s = \frac{3}{8 \pi^2}M_s^4 (x_s n_s( 2 x_s^2 + 1) -
                 \ln(x_s + n_s))
 \end{equation}
 where  $ x_s = k_s^f/M_s $  and  $ n_s = \sqrt{ 1 + x_s^2} $,
 $ k_s^f $ being the fermi momentum of the strange quarks.

 The total energy density of the quarks for the 3 flavour PC is
 given by
 \begin{equation}
 \epsilon_\rho = \epsilon_u  +  \epsilon_d + \epsilon_s + \frac{1}{2} F^2 q^2  
                 + \frac{\lambda_1^2}{4}( F^2 -( f_\pi^2))^2 
 \end{equation}
 From the effective potential given in \cite{ref6,ref23} for the $SU(3)$ case,
 there is an extra factor of 3/2 that multiplies the last term. This can be 
 absorbed, as we have done, by a redefinition, $ \lambda_1  =  A \lambda $, 
 where $A = \sqrt{3/2}$.

 The correction due to
 one gluon exchange interaction can also be incorporated in the evaluation
 of $\epsilon_u$, $\epsilon_d$ and $\epsilon_s$ above. We have discussed this
 in some detail in \cite{ref6}. In this paper we follow the prescription of
 Baym \cite{ref25}, as done in \cite{ref6}, to calculate the interaction 
 contribution to quark energy densities.

 \subsection {$\beta$-equilibrium in the PC phase}     
 We have the following general chemical potential relations for quark matter
 \begin{eqnarray}
      E^u_F &=& \mu_u \\
      E^d_F &=& \mu_d = \mu_s \\ 
      \mu_e &=& \mu_d - \mu_u \\ 
        n_e &=& \frac{\mu_e^3}{3\pi^2} 
 \end{eqnarray}
 The charge neutrality condition below further reduces the number of 
 independent chemical potentials to one.
 \begin{equation}
     \frac{2 n_u (\mu_u, q , F) - n_d (\mu_d, q ,F) - n_s (\mu_s)}{3} -n_e = 0 
 \end{equation}
 The baryon density is
 \begin{eqnarray}
     n_b &=& \frac{n_u (\mu_u, q , F) + n_d (\mu_d, q ,F) 
                      + n_s (\mu_s)}{3} \\
              n_s &=& (k_s^f)^3/(\pi^2)
 \end{eqnarray}
 For matter in $\beta$ equilibrium we need to add the electron energy density
 to the quark energy density above
 \[ \epsilon_e =  (1/4 \pi^2) \mu_e^4 \]
 The total energy density is
 \[ \epsilon  =  \epsilon_\rho + \epsilon_e \]
 The energy per baryon, $E_b = \epsilon / n_b$, then follows.

 For the pion condensed state, the ground state energy and the baryon density
depend on the variational parameters, the order parameter or the
expectation value, $F = \sqrt{ <\vec \pi>^2 + < \sigma>^2 }$ and the
condensate momentum, $|\vec q|$. To define the free energy at a fixed
baryon density then requires some care.

We go about this by defining a baryon chemical potential to go with a baryon 
density.
We have obtained both the baryon density and the energy density of the PC in
terms of the u,d,s quark fermi energies/chemical potentials.
First we construct the free energy
\begin{equation}
\Omega = \epsilon - n_b \mu_b
       =  \epsilon_\rho + \epsilon_e - n_b \mu_b
\end{equation}
The baryon chemical potential is defined as
\begin{equation}
     \mu_b  = \partial\epsilon/\partial n_b
\end{equation}
After meeting all the neutrality and equilibrium conditions above for fixed 
$F$ and $q$, we can write all the above variables as a function of a single 
variable, $\mu_u$.
We then minimize $\Omega$ independently with respect to $F$ and $q$. 
The $E_b$ etc then follow.

The results are presented in the tables below and in Figure 2a.
Fig 2b gives the the EOS for all the phases considered so far.
\begin{table}
\caption{Charge neutral, 3-flavour, beta-equilibrium pion condensed phase
         with $m_{\sigma} = 800$~MeV.  The columns are: u-quark chemical
         potential ($\mu_u$ in MeV), baryon density ($n_b$ in fm$^{-3}$),
         energy per baryon ($E_b$ in MeV), electron density
         ($n_e$ in fm$^{-3}$), ratio of densities of d-quark and u-quark
         ($n_d/n_u$), that of s-quark and u-quark ($n_s/n_u$), the order
         parameter ($F$ in MeV) and magnitude of the vector $q$.
         \label{tab:s800}}
\begin{center}
\begin{tabular}{cccccccc}
\hline
$\mu_u$ & $n_b$ & $E_b$ & $n_e$ & $n_d/n_u$ & $n_s/n_u$ & $F$ & $q$ \\
\hline
280.0 & 0.2972 &  984.94 & .2303E-02 & 1.916 & .0318 & 37.0406 & 2.5945\\
300.0 & 0.3645 &  981.48 & .1602E-02 & 1.731 & .2269 & 31.9655 & 2.6149\\
320.0 & 0.4591 &  994.07 & .1141E-02 & 1.599 & .3640 & 28.6471 & 2.9703\\
340.0 & 0.5409 & 1008.88 & .1173E-02 & 1.564 & .4004 & 30.4335 & 3.0216\\
360.0 & 0.6700 & 1043.21 & .9455E-03 & 1.499 & .4672 & 28.8195 & 3.6217\\
380.0 & 0.7628 & 1062.14 & .1063E-02 & 1.482 & .4847 & 31.6743 & 3.4574\\
400.0 & 0.8967 & 1104.94 & .4413E-03 & 1.345 & .6245 & 23.6066 & 3.9497\\
420.0 & 1.0472 & 1134.64 & .1283E-02 & 1.463 & .5040 & 36.1496 & 3.9839\\
440.0 & 1.1774 & 1168.76 & .4890E-03 & 1.317 & .6529 & 26.8871 & 4.0720\\
460.0 & 1.3225 & 1216.52 & .1794E-03 & 1.218 & .7529 & 19.3370 & 4.3125\\
480.0 & 1.5149 & 1246.21 & .3896E-03 & 1.267 & .7033 & 26.9453 & 4.3670\\
500.0 & 1.6900 & 1290.81 & .2212E-03 & 1.211 & .7597 & 23.0806 & 4.4865\\
\hline
\end{tabular}
\end{center}
\end{table}
\begin{figure}
\epsfig{file=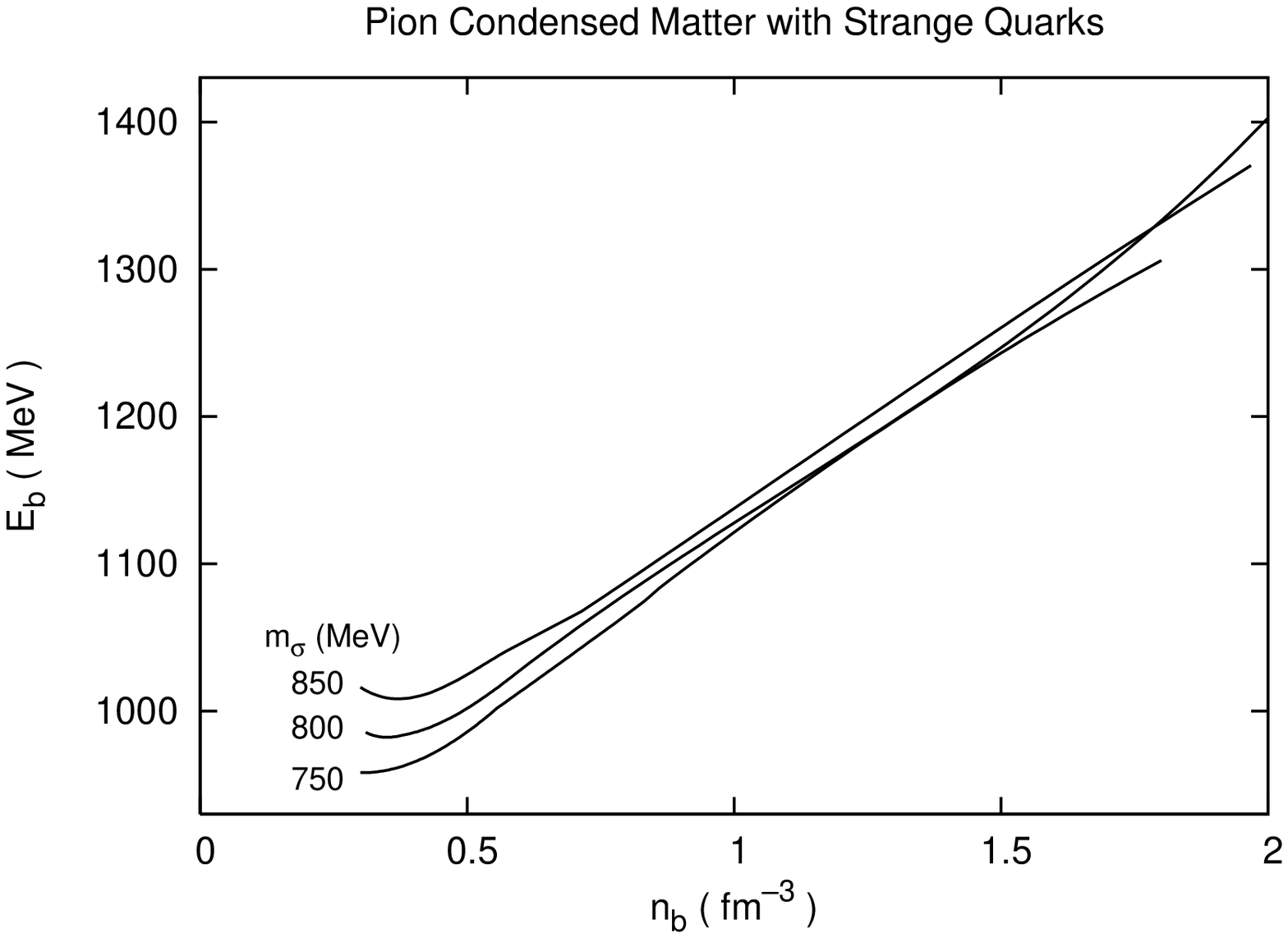,width=80mm}
\epsfig{file=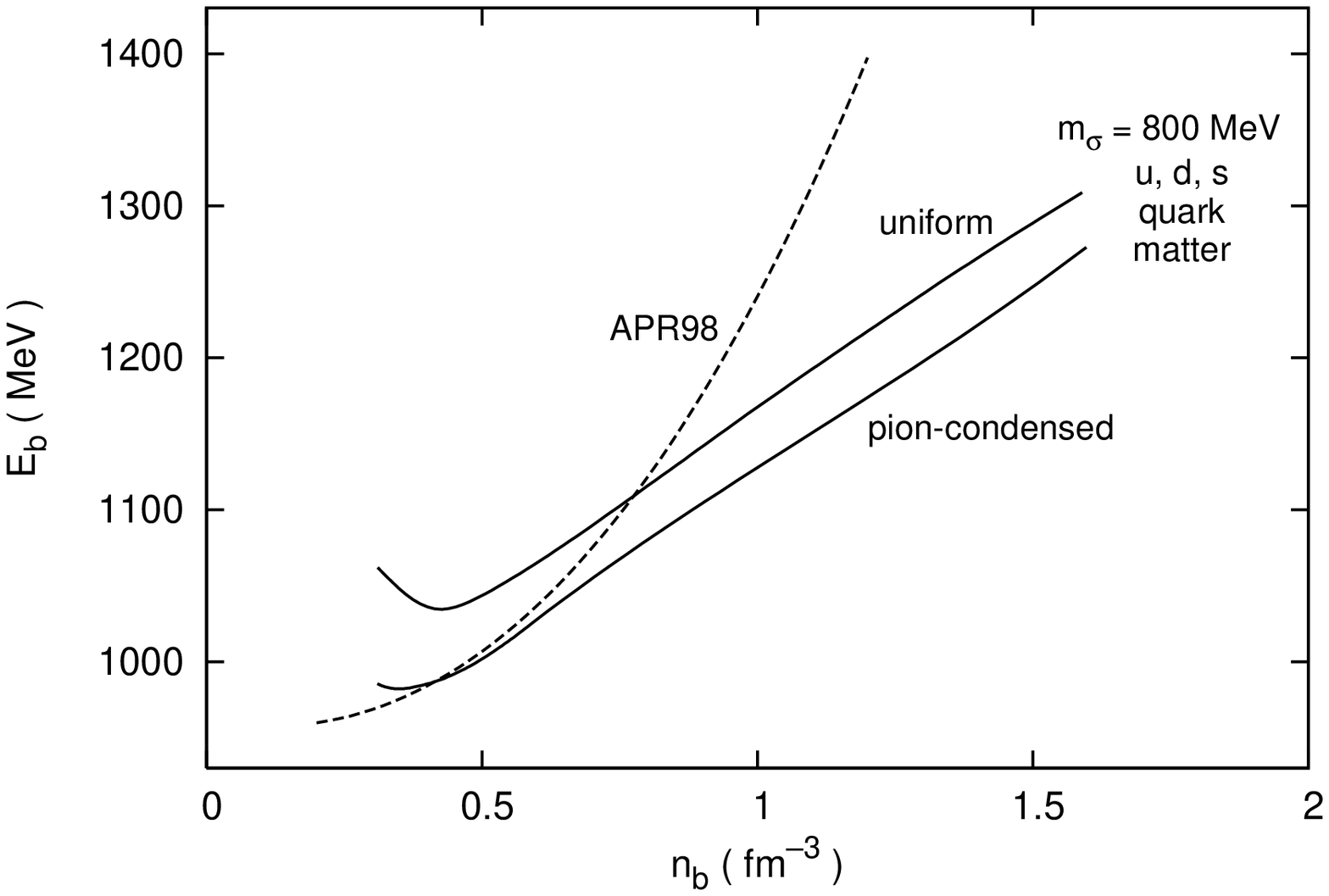,width=80mm}
\caption{(a) Upper panel: energy per baryon vs baryon number density for 3-flavour 
pion-condensed phase for three values of assumed tree-level mass of the
scalar meson $\sigma$. Charge neutrality and beta equilibrium are imposed. 
One-gluon exchange interaction is included using the prescription of 
Baym \cite{ref25}. (b) Lower panel: comparison of the equations of state of
the 3-flavour space uniform phase and pion-condensed phase for 
$m_{\sigma}=800$~MeV, with APR98 \cite{ref21}.}
\end{figure}

\section {The phase diagram and the equation of state}
The starting point for the phase diagram of QCD at finite density is as such:
At very low density we know that there is chiral SSB, with the pion as
the Goldstone boson - this breaks chiral symmetry spontaneously, leaving colour
symmetry unbroken. At very high density we have a colour SC pairing instability
for the quarks and of the many pairings investigated \cite{ref26} the CFL (colour flavour
locked) pairing is favoured -- this
spontaneously breaks colour and chiral symmetry. In between these limits the issue
of the ground state is open.

We approach this problem, from the low density end, by the effective lagrangian 
$L$ that can access both, nuclear and quark matter states, with or without chiral 
SSB. We have argued that this $L$ has validity upto energy scales of about 800 MeV.
The advantage of this $L$ is that it has only 3 couplings and no other free 
parameters. These can all be determined from experiment. 
However, at the mean field level, this $L$ is  good
for investigating chiral condensates but not colour condensates.

If we start at the high density end, as done by the several authors who
practice diquark
colour superconductivity, one knows the ground state at very high density,
when asymptotically free QCD makes calculation reliable.
However, for intermediate density  phenomenological four fermion interactions 
have to be introduced. This is equivalent to introducing a given value for the 
gap parameter, $\Delta$ \cite{ref26}. Also, another phenomenological input is
the bag parameter, $B$. Uncertainty attends these parameters.
It is not clear how this ground state transits into the low density chiral
SSB state.   

We start with saturation nuclear matter at nuclear density and this
persists till the nuclear matter gets squeezed into quark matter at
moderately higher density, when a pion condensed state takes over \cite{ref13}.
In all these phases, as we have found, chiral symmetry is spontaneously broken --
though the patterns of symmetry breaking keep changing with baryon density.
Since we have not investigated all other possible condensates, 
we cannot vouch for our neutral pion condensate being the best ground state.

It is, however, to be noted from the lecture notes of Baym \cite{ref25}, that
the neutral pion condensate is preferred over the charged pion condensate for
charge neutral nuclear matter, in the non relativistic limit (particularly
if we put the axial coupling constant $ g_A = 1 $). We have found that this 
is the case for charge neutral quark matter as well \cite{ref27}.
It is worth pointing out that all these states that have lower
energy than the chirally restored CRQM state, are chiral symmetry broken
states.

At even higher density the most likely state is a diquark condensate - a
colour flavour locked (CFL) \cite{ref23} state - which can persist till
arbitrarily high density. Such a ground state spontaneously breaks both 
chiral and colour symmetry.
Diquark condensates are unlikely at moderately high density as they depend
on the quark density of states and so the pion condensate is most likely
at such densities.

There is an important issue that arises here and that is the comparison
between the diquark condensate state and the pion condensed state.
In this case the starting point is a chiral symmetric four fermion
interaction which can accommodate both chiral (quark-antiquark colour
singlet) condensates and diquark condensates. Such a comparison has 
already been done by Sadzikowski \cite{ref28} in the context of a NJL 
chiral symmetric model, for the case of 2 flavours -- $ SU(2)_L \times
SU(2)_R $. What is done is at the level of mean field theory. The NJL 
model has four fermion interactions in terms of the
quark bilinears corresponding to the $\sigma$ and $\pi$ field quantum
numbers, with a common dimensional coupling, G. If we are interested in
a ground state carrying sigma and/or pion condensates we can replace these
quark bilinears by the corresponding $\sigma$ and $\pi$ EV's in
the MFT. This yields the ground state energy of the space uniform SSB
and the PC states.
Alternatively, this NJL can be mapped to our linear sigma model and
the ground state thereof, which has been considered earlier.

To get to the diquark condensate state  we have to Fierz
transform this NJL chiral, $L$, and look for its projection into the diquark
condensate channel and follow the same procedure of MFT. This projection
gives the diquark condensate lagrangian used by \cite{ref26,ref28}, with a 
four fermion coupling, $G^{'}=G/4$ \cite{ref28}.

Working with simultaneous MF condensates, corresponding to  space uniform 
    chiral SSB (which generates  a spontaneous mass for the quarks (no PC)) and 
    diquark SC (which gives rise to colour SC), they \cite{ref28} find that for
    the above value of the two $G^{'}$s, that follow from the Fierz projection, 
    the chiral condensate is always the preferred state up till the limit of 
    validity of this model.

    However, they find that with arbitrary (higher) values of $G^{'}$, it is 
    possible to have a phase transition from a space uniform chiral SSB to a 
    diquark condensate at some large density.
    In particular, they consider,  $G^{'}= G$, and find that from a pure chiral
    SSB state, with $ m = g_yf_\pi$ = 301~MeV, there is a continuous phase 
    transition to a mixed chiral SSB with a small admixture of diquark 
    condensate state at a quark 
    chemical potential of about 0.33 GeV. For the higher value of, $m = 500$~MeV 
    we employ, this is expected to occur at a higher value of $\mu$. This is 
    followed by a first order phase transition at slightly
    higher $\mu$, to also a dominantly diquark condensate mixed state.

    This state then evolves to an almost entirely diquark condensate.   

Till now we have not cosidered the PC state. The PC state always has lower
    energy (free energy) than the uniform chiral SSB state. It would then be 
    reasonable to expect that the PC state and not the uniform state would be 
    the preferred state of chiral SSB. Since chiral SSB persists till much higher 
    density ($\mu$) in the PC state we expect that with the inclusion of this state 
    the transition to the colour SC state will be pushed to higher density 
    ($\mu$).

A following work by the same author addresses this matter by considering 
simultaneous MF condensates of the, PC, and the diquark condensate \cite{ref29}.
It is found that the first order transition does shift to higher density 
($\mu = 400$~MeV).  This is considered only for the particular case  
$ G^{'} = G/2$, and for $m = 301$~MeV. 
For our larger value of $m$ this is expected to occur at larger $\mu$. 
Furthermore, in this case it seems after the first order transition occurs, the 
chiral SSB order parameter, $M$, goes to zero, indicating that this state 
has no chiral SSB but only colour SSB.
Of course, this is so as these works deal with the two-flavour case -- where 
the colour diquark condensate is a chiral singlet.

Realistically, we must consider 3 flavours, since the quark chemical potential 
is much greater than the strange quark mass. In this case we are very likely 
to have the CFL state as the lowest energy state. The criterion for this is 
given in \cite{ref30} and is  $\Delta > {m_s^2}/{4\mu}$, which is easily 
satisfied.  Furthermore, in this case the diquark condensate is a 
colour-flavour condensate which has both chiral SSB and colour SSB, albeit in 
a manner different to the PC. 

The deciding question is then what effective $L$ is to be used. There is no
derivation of the exact effective $L$ from QCD at arbitrary scales. There are
many four fermion type (or higher order) interactions derived or rather
motivated from various points of view, e.g.\ instantons, gluon exchange etc.
However, there is no unique $L$ that is derived as such. It is therefore
necessary to have an argument to decide this issue. See also \cite{ref34}
for a related discussion.

 Since the NJL chiral $L$ identifies with our linear sigma model, and this
 latter model is what we have used as a valid model till centre of mass
 energies/scales of less than 800 MeV, the right procedure would be to take this
 model to describe physics upto this scale. In this case, as we have argued,
 the PC is the preferred ground state, till the scale of validity of our
 effective $L$ (this is for $ G^{'} =  G/4 $). Even if we relax this to
 larger $G^{'}$  we have a three-flavour PC, till $\mu$ well above 400 MeV, 
 followed by a CFL state, with increasing density. As the tables suggest such 
 values of, $\mu$, correspond to baryon density 5--6 times nuclear density.
 This makes the PC as likely state in neutron star cores. 

 With increasing baryon density we then find the following hierarchy. At
 nuclear density and above we have nuclear matter with chiral SSB, followed
 by the pion condensed quark matter, again with chiral SSB, albeit with a
 different realization and finally a transition to the diquark CFL state
 which also has chiral SSB (and colour SSB), with yet another
 realization. A point to note is that CRQM or free fermi seas are
 unstable. To one's surprise at zero temperature, at any finite density
 chiral symmetry is never restored!  A similar result has also been
 obtained in 1+1 dimension \cite{ref35}.

\section{Discussion and implications for Neutron/Strange stars}

\subsection{Stars}
 We have used the quark based linear sigma model for this analysis,
 where the tree level sigma mass was set to be around 800 MeV as indicated
 by the analysis of Schechter et al \cite{ref14},
 by matching the results for meson-meson scattering from this model, with
 experiment.

   Such a tree level sigma mass also ruled against conventional SQM being the
   ground state of matter \cite{ref6}.
   Strange stars require the absolute stability of SQM. Since this has been
   shown to be highly implausible so is then the existence of strange stars.

   Independently, we find (Bhattacharya and Soni, in preparation) that 
   realistic Neutron stars with PC cores occur only when the tree level sigma 
   mass in this model is in a small window, 750 - 850 MeV.

This can be seen from $E_b$ vs $1/n_b$ diagram for the phases
we have considered (Fig 3). The negative slope in this figure gives the pressure 
and the intercept on the vertical axis the baryon chemical potential. The Maxwell 
contruction of a common tangent to these curves gives the pressure and 
baryon chemical potential at which the transition occurs with a density 
discontinuity.

We find that at  $m_\sigma = 800$~MeV we get a transition with a relatively 
small pressure and a small density discontinuity. This permits us to have a 
well
developed PC core in a neutron star. The small density jump, besides, reassures
us that a mixed \cite{ref31} phase would give results that would be very 
similar. Of course, with the APR98 \cite{ref21} EOS that we use, we do not 
have analytical expressions for all variables, which precludes a mixed phase 
calculation.

At  $m_\sigma  = 850$~MeV, the transition moves to higher density and 
higher pressure with a larger density jump, so much so that there there is no 
PC core for the star -- as the maximum mass instability for the star occurs
before a core can form. 

On the other side, at $m_\sigma = 750$~MeV, the curves for the two phases
do not permit a common tangent construction any more -- so again a star with a 
PC core is ruled out.
\begin{figure}
\epsfig{file=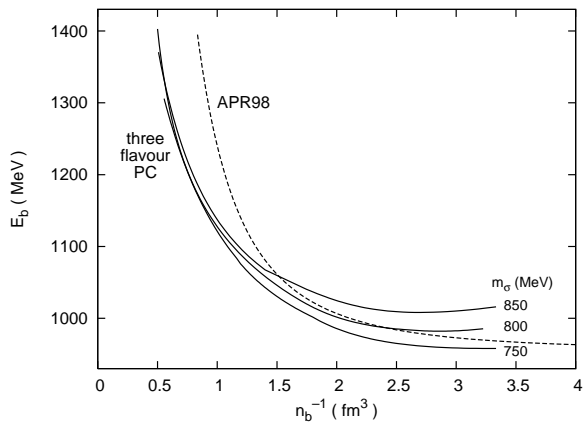,width=80mm}
\caption{The Maxwell construction: Energy per baryon plotted against the 
reciprocal of the baryon number density for APR98 equation of state (dashed line) 
and the 3-flavour pion-condensed (PC) phase, for three different values of 
$m_{\sigma}$ (solid lines).
A common tangent between the PC phase and the APR98 phase in this diagram
gives the phase transition between them.  The slope of a tangent gives the
negative of the pressure at that point, and its intercept gives the chemical 
potential. As this figure indicates, the transition pressure moves up with
increasing $m_{\sigma}$, and at $m_{\sigma}$ below $\sim 750$~MeV a common
tangent between these two phases cannot be obtained.}
\end{figure}
Is this a coincidence that a single parameter in our effective $L$, the mass of
the sigma or $\lambda$, plays a crucial role? Is it fortuitous that the tree level
sigma mass set by scattering experiments sits in a small window that simultaneously
rules out SQM as the absolute ground state of matter and also can provide us with 
neutron stars that can have pulsar range magnetic PC cores?

The problem in sustaining a PC core with a nuclear exterior is that we have a 
stiff exterior with a soft interior -- a rather unstable situation. It is thus 
not so surprising that very particular conditions must obtain for this to occur. 

\subsection{The Effective Lagrangian and Mean Field Theory}
   The question of the validity of our effective $L$ is of the essence.
Furthermore, since we do not go any further than MFT, the question of the
validity of MFT is another equally important question. Linked to this is the
question of QCD corrections.

\vspace*{1ex}
\noindent
\textit{1. Meson sector}\\
Let us begin with the part of the Lagrangian that describes the meson sector.
As we have said earlier the justification for this, its range of validity and
the value of the coupling $\lambda$ or the sigma mass have been very clearly 
spelt out in \cite{ref14}. It will be useful to review this.
As stated in this work it is the tree level Lagrangian that is used to describe 
$\pi$-$\pi$ scattering in the scalar channel. This tree-level amplitude is then
improved by unitarising the K matrix. A very good fit to scattering data in this 
channel follows if we choose the sigma mass in the tree level $L$ to be 
$800$--$850$~MeV.  The validity of this vis a vis the data is upto 
$\sqrt{s} < 800$~MeV.

The actual mass and width of the sigma can then be gleaned by looking at the 
pole in this K unitarized amplitude and is found to be $460$--$600$~MeV.

We have used this to fix the tree level coupling/sigma mass for our $L$. Since 
we work only in MFT (tree level), this also perhaps tells us that we may be off 
the mark by about 30\%.
  
Another cautionary point is that the above results are based only on scalar channel scattering and may not apply generally. It is also possible to fit the data
with chiral perturbation theory (infinite $ m_\sigma $) and the $\rho$
resonance. 
It is possible that by excluding the $\rho$ we may be missing some short-distance
repulsion, which might result in making our PC EOS too soft.

\vspace*{1ex}
\noindent
\textit{2. Quark sector}\\
For obvious reasons one cannot find an analagous scattering experiment for the
quark meson sector. However, there is other indirect evidence in this sector.
\begin{enumerate}
\item[i)] The quark soliton nucleon that we have used to fix the coupling, $g_y$, 
   satisfies the Goldberger Triemann relation exactly \cite{ref32} in accord 
   with chiral symmetry. As stated in the introduction several independent 
   properties of this nucleon, for example, magnetic moments compare well with 
   experiment.
\item[ii)] In the large $N_c$ expansion, the soliton (and other physics) 
   receives $1/N_c$ corrections from loops.
\item[iii)] Goksch \cite{ref12} finds that the finite temperature screening 
   masses of mesons computed in lattice QCD compare very favorably with those 
   calculated using finite temperature MFT for precisely our $L$ (without 
   including gluons).  However, the couplings used are slightly different 
   to ours,  $g_y = 3.3$ and sigma mass of 600~MeV.
\end{enumerate}
Georgi and Manohar \cite{ref3} use a non linear sigma model version of the 
above $L$ to do effective cutoff field theory and argue for larger energy 
scale of chiral symmetry breaking vis a vis confinement. Using this scale 
for the cutoff, they are able to naively but consistently include arbitrary 
meson/quark loops. Further, they heuristically argue that the QCD coupling 
is weak for the chiral SSB vacuum and thus gluon loops may be ignored. 

Actually, as we have stated in the introduction, QCD can have multiple scales
\cite{ref5} - a confinement scale, a chiral symmetry restoration scale and a 
compositeness scale for the pion. We would like to point out that the arguments
of \cite{ref3} would go through if the compositeness scale was substituted
for the chiral restoration scale, since this scale is used for a momentum
cutoff.  This will allow the chiral restoration scale to fall anywhere in
between the confinement and compositeness scale -- a likely possibility suggested 
by the lattice.

Parenthetically, we remark that the non-perturbative input of a finite
mass sigma particle/resonance may do the same job. This is also supported by 
observations in Ref.\ \cite{ref14} and references therein. This also supports 
our effective $L$, which is built on the premise that the chiral SSB energy 
scale is larger than than the confinement scale. 

\vspace*{1ex}
\noindent
\textit{3. QCD sector}\\
We have no evident justification for neglecting this strong sector or at most 
doing simply one gluon exchange corrections.

Ref.~\cite{ref3} makes the heuristic argument that the gluon coupling becomes 
strong to precipitate chiral SSB, after which it is expected to be perturbative 
in the new broken vacuum. They also argue, that from the colour and 
electromagnetic hyperfine splittings in the baryon sector of the non 
relativistic quark model one may infer an $\alpha_s = 0.28$.

At finite density, there is screening in quark matter. The correct expansion 
parameter is then the screened charge and not the usual running 
coupling constant written as a function of the quark chemical potential which is
commonly used. Such a screened charge has been formally constructed by one of us
\cite{ref33} and is clearly much smaller than the usual running coupling.
This makes a perturbative expansion in this coupling more plausible for dense 
quark matter.

Further, the ground state of quark matter, the pion condensate, has spin-charge 
ordering. Long range forces like one gluon exchange then inhibit any change
in the ground state and stabilize it against fluctuations as in many condensed 
matter situations where the coulomb forces are operative.  In this case 
MFT may be valid.

All the calculations referred to above for quark matter states have been carried 
out in MFT. We have tried to give some justification for its use. However,
we cannot provide any rigorous proof for the validity of MFT. It is also the
simplest thing to do and works well in many cases even if the coupling is strong.

Since we began this work with an eye to neutron stars it may be appropriate
to present our finding. We expect that the density profile of the star will start 
with the nucleonic EOS on the outside and go to a pion condensate in the interior 
and could well go to a colour flavor locked state at the centre if the density 
there is large enough; though, this is unlikely if we have a pion condensate 
as its softness may result in a smaller maximum mass.

\subsection*{Acknowledgements}
This work began from a weekly series of lectures on the origin of magnetic fields
in Astrophysics organised by Siraj Hasan at the Indian Institute of Astrophysics
in which, besides us, Pijush Bhattacharjee, C. Sivaram and C.S. Shukre participated.
Thanks go to them all.  We thank W. Broniowski for sharing some of their computer
codes with us.  VS thanks Prof. J. Schechter for reading the manuscript,
D. Sahadev, C. S. Aulakh and Nirvikar Prasad for discussions, and the Raman 
Research Institute for ample and steady hospitality.

\end{document}